\documentclass[a4paper,11pt]{article}

\usepackage[colorlinks=true,allcolors=blue]{hyperref}

\usepackage{amsmath,amssymb,mathtools}
\usepackage{color}
\usepackage{graphicx}
\usepackage{subfigure}
\usepackage{cite}
\usepackage{hyperref}
\usepackage{multirow,makecell}
\usepackage{textcomp}
\usepackage{wasysym}
\usepackage{setspace}
\usepackage{verbatim}
\usepackage{float}
\usepackage{ulem}
\usepackage[utf8]{inputenc}
\usepackage{array}
\usepackage{booktabs}

\newcolumntype{C}[1]{>{\centering\arraybackslash}m{#1}}
\graphicspath{{figures/}}

\usepackage[text={17.2cm,25.2cm},centering]{geometry}

\numberwithin{equation}{section}
\renewcommand\arraystretch{1.4}

\begin{document}

\title{Holographic spectral functions of exotic spin-1 mesons}

	\author{Yan-Qing Zhao$^{1}$\footnote{zhaoyanqing@hainnu.edu.cn} and Zhou-Run Zhu$^{2}$\footnote{zhuzhourun@zknu.edu.cn}}

	\maketitle
    \date{}
	\vspace{-10mm}
	
	\begin{center}
		{\it
			$^{1}$ College of Physics and Electronic Engineering, Hainan Normal University,\\ Haikou 571158, China\\
            $^{2}$ School of Physics and Telecommunications Engineering, Zhoukou Normal University,\\ Zhoukou 466001, China\\\vspace{1mm}
		}
		\vspace{10mm}
	\end{center}

\begin{abstract}
We develop a general holographic framework for computing thermal spectral functions of exotic spin-1 mesons. A bulk Proca mass encodes the canonical ultraviolet dimension of the interpolating operator, allowing channels with different ultraviolet scaling to be treated within a unified membrane-flow formalism in the zero-spatial-momentum limit. We test this framework in a soft-wall model with a gluon-condensate background for the hybrid $\pi_1$ and tetraquark-like $Z_c$ channels. The results show that dissociation temperatures of both channels lie below the phase transition temperatures. Increasing the gluon-condensate parameter $c$ delays dissociation in both channels, but through different effects: both the vacuum-mass shift and thermal spectral deformation contribute in the $\pi_1$ channel, whereas the latter dominates in the $Z_c$ channel. Nevertheless, for the same environment, the $Z_c$ resonance dissociates at a lower temperature than the $\pi_1$ resonance.
\end{abstract}

\maketitle

\newpage{}

\tableofcontents

\newpage{}

\section{Introduction}\label{sec:01_intro}

The properties of strongly interacting matter under extreme conditions are among the central topics in high-energy nuclear physics. Relativistic heavy-ion collision experiments at RHIC and the LHC provide an important window into the quark-gluon plasma (QGP), where the behavior of strongly coupled QCD matter can be explored experimentally~\cite{BRAHMS:2004adc,PHENIX:2004vcz,PHOBOS:2004zne,STAR:2005gfr}. In-medium modifications of hadrons, including mass shifts, thermal broadening, and dissociation, provide valuable information about the interaction between bound states and the strongly coupled medium. Heavy quarkonia have traditionally
been considered important probes of deconfinement due to their sensitivity to the thermal environment~\cite{Matsui:1986dk,Satz:2005hx}. Recently, increasing attention has been
devoted to exotic hadrons, such as tetraquark and hybrid meson candidates, whose unconventional multiquark and gluonic structures offer new possibilities for investigating nonperturbative aspects of QCD~\cite{Lebed:2016hpi,Guo:2017jvc,Meyer:2015eta,BESIII:2015cld}.

Understanding the in-medium behavior of exotic hadrons from first principles remains a challenging problem because of the strongly coupled nature of QCD in the relevant energy regime. The AdS/CFT correspondence~\cite{Maldacena:1997re,Gubser:1998bc,Witten:1998qj} and its bottom-up holographic extensions provide a useful nonperturbative framework for studying QCD theories~\cite{Gubser:2008ny,DeWolfe:2010he,He:2013qq,Yang:2014bqa,Dudal:2017max,Critelli:2017oub,Grefa:2021qvt,Arefeva:2020vae,Zhao:2023gur,Zhao:2022uxc,Tong:2026dwb,Cai:2024eqa,Fu:2024nmw,He:2023ado,Li:2023mpv,Cai:2022omk,Li:2026pkd}.
Although holographic QCD models are not derived directly from QCD, they have successfully described various hadronic properties, including mass spectra, decay constants, and finite-temperature spectral functions~\cite{Erlich:2005qh,Karch:2006pv,Branz:2010ub,deTeramond:2005su,Sakai:2004cn,Sakai:2005yt,Zhang:2026zoz}. In particular, soft-wall models with generalized channel-dependent profiles have been widely used to reproduce nonlinear Regge trajectories of heavy and exotic states~\cite{Braga:2016wkm,Braga:2015jca,Braga:2017bml,MartinContreras:2020cyg,Fujita:2009ca}.

Spectral functions obtained from holographic retarded correlators provide a direct characterization of the in-medium properties of hadrons. The resonance peaks encode information about thermal masses and widths, allowing one to study the survival and dissociation of bound states in a hot medium. Previous holographic studies have investigated the spectral functions of conventional quarkonia as well as exotic tetraquark and hybrid states under different extreme conditions, including finite temperature and density~\cite{Bellantuono:2014lra,Braga:2017oqw,Braga:2018zlu,Toniato:2025gts,Zhu:2026pqm,Ballon-Bayona:2024twa}, magnetic fields~\cite{Zhao:2021ogc,Dudal:2014jfa,Zhao:2023pne,Jena:2024cqs}, rotation~\cite{Zhao:2023pne,Braga:2023fac,Zhu:2024uwu}, and anisotropic backgrounds~\cite{Chang:2024ksq}. However, a systematic framework for constructing spectral functions of exotic spin-1 mesons with explicit bulk mass contributions remains less explored. Such a framework is essential for describing different exotic channels within a unified holographic setup, since the bulk mass is directly related to the scaling dimension of the dual operator and distinguishes different interpolating structures.

In addition to thermal effects, the nonperturbative structure of the QCD vacuum can also influence the properties of hadronic states. The gluon condensate, as an important characterization of the QCD vacuum, encodes nonperturbative gluonic contributions and has been incorporated into holographic models through dilaton black hole backgrounds~\cite{Brown:2006vn,Kim:2008ax,Zhu:2021vkj,Afonin:2020crk}. These backgrounds provide a phenomenological approach to studying how vacuum properties modify hadron spectra and the confinement-deconfinement transition. Previous studies have shown that the gluon condensate can affect the stability and binding properties of heavy quarkonium~\cite{Tahery:2022pzn,Zhao:2019tjq,Zhang:2020upv}, suggesting that the gluonic structure of the vacuum may also play an important role in the in-medium evolution of
exotic hadrons. Nevertheless, the influence of the gluon-condensate background on the spectral functions and thermal dissociation of exotic spin-1 mesons has not been systematically investigated.

In this work, we develop a general holographic framework for calculating spectral functions of exotic spin-1 mesons by including the bulk mass contribution of the five-dimensional spin--1 field. This framework allows different exotic channels to be described through channel-dependent bulk parameters and background profiles. As a concrete application, we study the hybrid meson $\pi_1$ and the tetraquark-like $Z_c$ channel in a holographic background with a gluon condensate. We analyze how the gluon-condensate
deformation modifies their vacuum spectra and finite-temperature spectral functions, and investigate the corresponding thermal evolution and dissociation behavior. Our results provide a unified holographic description of exotic spin-1 mesons and offer insight into the role of nonperturbative vacuum structures in their in-medium properties.

The paper is organized as follows. In section~\ref{sec:spectral-membrane}, we derive the membrane-flow equations for massive spin--1 fluctuations and discuss the holographic normalization of the spectral functions. In section~\ref{sec:exotic-gluon-condensate}, we apply this framework to the $\pi_1$ and $Z_c$ channels in the gluon-condensate background and analyze their vacuum spectra and finite-temperature spectral functions. Section~\ref{sec:summary} contains our summary and discussion.

\section{Spectral functions from membrane flow at zero spatial momentum}
\label{sec:spectral-membrane}

In this section, we derive the retarded two-point function and the corresponding spectral density for an effective massive spin--$1$ fluctuation in a finite-temperature soft-wall background. Throughout the analysis, we consider the exotic meson at rest with respect to the thermal medium and set its spatial momentum to zero, ${\bf{q}}=0$, and therefore no distinction between transverse and longitudinal spatial polarizations is required. In a spatially isotropic background, the three spatial polarizations are degenerate and are governed by a single radial flow equation.

We consider a general diagonal black-hole geometry,
\begin{equation}
  ds^2=-g_{tt}(z)\,dt^2
  +\sum_{i=1}^{3}g_{x_ix_i}(z)\,dx_i^2
  +g_{zz}(z)\,dz^2,
  \label{eq:diag-metric}
\end{equation}
where all metric functions appearing on the right-hand side are positive outside the horizon. The 4--dimensional QCD boundary is located at $z=0$, while the non-extremal horizon lies at $z=z_h$. The effective vector field dual to the spin--$1$ interpolating operator in channel $X$ is described by the Proca action
\begin{equation}
  S_V=-\frac{1}{4g_5^2}
  \int d^5x\,\sqrt{-G}\,e^{-\Phi_X(z)}
  \left(F_{mn}F^{mn}+2m_{5,X}^2V_mV^m\right),
  \qquad
  F_{mn}=\partial_mV_n-\partial_nV_m.
  \label{eq:proca-action}
\end{equation}
Here $\Phi_X(z)$ is the channel-dependent soft-wall profile. In the present bottom-up construction, $\Phi_X(z)$ controls the infrared dynamics, whereas the bulk mass fixes the leading ultraviolet scaling of the dual operator. For a one-form field in AdS$_5$, the mass-dimension relation is
\begin{equation}
  m_{5,X}^2L^2=(\Delta_X-1)(\Delta_X-3).
  \label{eq:vector-mass-dim}
\end{equation}
Accordingly, $\Delta_X=5$ gives $m_{5,X}^2L^2=8$, while $\Delta_X=6$ gives $m_{5,X}^2L^2=15$. In the phenomenological implementation used here, $\Delta_X$ is identified with the canonical ultraviolet dimension of a representative local QCD interpolating current. The currents adopted for the $\pi_1$ and $Z_c$ channels are specified in
section~\ref{subsec:fixed-channel-profiles}. Possible anomalous dimensions and operator mixing are neglected.

Varying Eq.~\eqref{eq:proca-action} gives
\begin{equation}
  \partial_m\!\left(\sqrt{-G}\,e^{-\Phi_X}F^{mn}\right)
  -\sqrt{-G}\,e^{-\Phi_X}m_{5,X}^2V^n=0.
  \label{eq:proca-eom}
\end{equation}
For $m_{5,X}^2\neq0$, taking the divergence of this equation yields the Proca constraint
\begin{equation}
  \partial_n\!\left(
  \sqrt{-G}\,e^{-\Phi_X}m_{5,X}^2V^n
  \right)=0.
  \label{eq:proca-constraint}
\end{equation}
In a generic massive Proca theory, $V_z$ is not a gauge degree of freedom and therefore cannot be set to zero arbitrarily by a gauge choice. However, at the zero spatial momentum considered in this work, the spatial components $V_i$ relevant to the spin--$1$ spectral function decouple exactly from the temporal--radial sector $(V_t,V_z)$. Consequently, restricting to $V_i$ while setting $V_t=V_z=0$ constitutes a consistent truncation allowed by the equations of motion, rather than a choice of radial gauge. For a single spatial polarization, we take
\begin{equation}
  V_m(t,z)=e^{-i\omega t}V_m(z),
  \qquad
  V_t=V_z=0,
  \qquad
  V_i\neq0,
  \qquad
  V_{j\neq i}=0.
  \label{eq:zero-momentum-spatial-ansatz}
\end{equation}
The equation of motion for $V_i$ then takes the form
\begin{equation}
  \partial_z\!\left[\mathcal K_i(z)\,\partial_zV_i\right]
  +\sqrt{-G}\,e^{-\Phi_X}g^{x_ix_i}
  \left(\frac{\omega^2}{g_{tt}}-m_{5,X}^2\right)V_i=0.
  \label{eq:zero-momentum-vector-eom}
\end{equation}
with
\begin{equation}
  \mathcal K_i(z)
  =
  \sqrt{-G}\,e^{-\Phi_X(z)}g^{zz}(z)g^{x_ix_i}(z).
  \label{eq:Ki-def}
\end{equation}
In a spatially isotropic geometry,
$g_{x_1x_1}=g_{x_2x_2}=g_{x_3x_3}\equiv g_{xx}$, this equation is the same for all three spatial polarizations.

\subsection{Radial flow of the spatial response}
\label{subsec:zero-momentum-flow}

We introduce the electric field
\begin{equation}
  E_i(z)\equiv F_{x_it}=i\omega V_i(z)
  \label{eq:Ei-def}
\end{equation}
and the UV-oriented radial response current
\begin{equation}
  \mathcal J_i(z)
  \equiv
  \frac{1}{g_5^2}
  \sqrt{-G}\,e^{-\Phi_X(z)}F^{zx_i}
  =
  \frac{1}{g_5^2}\mathcal K_i(z)\,\partial_zV_i(z).
  \label{eq:radial-current}
\end{equation}
The sign in Eq.~\eqref{eq:radial-current} is adapted to a radial coordinate that increases from the 4-dimensional QCD boundary toward the horizon. With this orientation, the retarded correlator below is obtained from $-\mathcal J_i/V_i$.

The radial conductivity associated with the spatial polarization $i$ is
\begin{equation}
  \sigma_i(\omega;z)
  \equiv
  \frac{\mathcal J_i(\omega;z)}{E_i(\omega;z)}.
  \label{eq:sigma-zero-momentum-def}
\end{equation}
Using Eq.~\eqref{eq:zero-momentum-vector-eom}, one finds
\begin{equation}
  \frac{\partial_z\mathcal J_i}{E_i}
  =
  i\omega
  \sqrt{\frac{g_{zz}}{g_{tt}}}\,
  \Sigma_i(z)
  \left(1-\frac{m_{5,X}^2g_{tt}}{\omega^2}\right).
  \label{eq:partial-current-zero-momentum}
\end{equation}
with
\begin{equation}
  \Sigma_i(z)
  =
  \frac{1}{g_5^2}e^{-\Phi_X(z)}
  \sqrt{\frac{-G}{g_{zz}(z)g_{tt}(z)}}\,
  g^{x_ix_i}(z).
  \label{eq:sigma-i-def}
\end{equation}
For an isotropic background, all $\Sigma_i$ coincide, and we denote their common value by $\Sigma(z)$. Moreover, the Bianchi identity gives
\begin{equation}
  \frac{\partial_zE_i}{E_i}
  =
  i\omega
  \sqrt{\frac{g_{zz}}{g_{tt}}}\,
  \frac{\sigma_i}{\Sigma_i}.
  \label{eq:partial-electric-zero-momentum}
\end{equation}
Combining Eqs.~\eqref{eq:partial-current-zero-momentum} and \eqref{eq:partial-electric-zero-momentum}, we obtain the zero-momentum Riccati flow equation
\begin{equation}
  \partial_z\sigma_i
  =
  -i\omega
  \sqrt{\frac{g_{zz}}{g_{tt}}}
  \left[
  \frac{\sigma_i^2}{\Sigma_i}
  -\Sigma_i
  \left(1-\frac{m_{5,X}^2g_{tt}}{\omega^2}\right)
  \right].
  \label{eq:flow-zero-momentum}
\end{equation}
No longitudinal auxiliary variables are required: at ${\bf{q}}=0$, each spatial polarization is governed by the single closed equation \eqref{eq:flow-zero-momentum}. In the massless limit, the standard Maxwell flow equation is recovered.

It is advantageous to reformulate the flow problem in terms of the radial response function,
\begin{equation}
  \mathcal G_i(\omega;z)
  \equiv
  -\frac{\mathcal J_i(\omega;z)}{V_i(\omega;z)}
  =-i\omega\,\sigma_i(\omega;z).
  \label{eq:radial-response-def}
\end{equation}
Its flow equation is
\begin{equation}
  \partial_z\mathcal G_i
  =
  \sqrt{\frac{g_{zz}}{g_{tt}}}
  \left[
  \frac{\mathcal G_i^2}{\Sigma_i}
  +\Sigma_i\left(\omega^2-m_{5,X}^2g_{tt}\right)
  \right].
  \label{eq:response-flow-zero-momentum}
\end{equation}
Equation~\eqref{eq:response-flow-zero-momentum} is algebraically equivalent to Eq.~\eqref{eq:flow-zero-momentum}, but contains no explicit inverse powers of $\omega$ and is therefore generally more stable for numerical integration in the low-frequency regime.

\subsection{Infalling condition at the horizon}
\label{subsec:zero-momentum-horizon}

We assume a non-extremal horizon with
\begin{equation}
  g_{tt}(z)
  =g_{tt}^{(1)}(z_h-z)+\mathcal O((z_h-z)^2),
  \qquad
  g_{zz}(z)
  =\frac{g_{zz}^{(-1)}}{(z_h-z)}+\mathcal O(1).
  \label{eq:near-horizon-metric}
\end{equation}
The Hawking temperature is determined by
\begin{equation}
  \sqrt{\frac{g_{tt}^{(1)}}{g_{zz}^{(-1)}}}
  =4\pi T,
  \label{eq:kappa-horizon}
\end{equation}
so that
\begin{equation}
  \sqrt{\frac{g_{zz}}{g_{tt}}}
  =\frac{1}{4\pi T(z_h-z)}+\mathcal O(1),
  \qquad
  \sqrt{\frac{g_{tt}}{g_{zz}}}
  =4\pi T(z_h-z)+\mathcal O((z_h-z)^2).
  \label{eq:near-horizon-ratios}
\end{equation}

For the Fourier convention $e^{-i\omega t}$, the retarded solution is selected by the infalling condition
\begin{equation}
  V_i(z)
  =
  (z_h-z)^{-i\omega/(4\pi T)}
  \left[v_{i,h}+\mathcal O((z_h-z))\right].
  \label{eq:incoming-vector-zero-momentum}
\end{equation}
Using Eqs.~\eqref{eq:radial-current}, \eqref{eq:sigma-i-def}, and \eqref{eq:incoming-vector-zero-momentum}, one obtains
\begin{equation}
  \mathcal J_i
  =\Sigma_i(z_h)E_i+\mathcal O((z_h-z) E_i).
  \label{eq:incoming-current-relation}
\end{equation}
Therefore, the regular horizon condition for the conductivity is
\begin{equation}
  \sigma_i(\omega;z_h)=\Sigma_i(z_h).
  \label{eq:horizon-conductivity-zero-momentum}
\end{equation}
Equivalently, the response function satisfies
\begin{equation}
  \mathcal G_i(\omega;z_h)
  =-i\omega\,\Sigma_i(z_h).
  \label{eq:horizon-response-zero-momentum}
\end{equation}
The Proca mass does not modify the leading horizon value because $g_{tt}(z_h)=0$; its effect enters through the radial evolution away from the horizon.

In numerical integrations, the flow is initialized at $z_*=z_h-\epsilon_h$, with $\epsilon_h\ll z_h$, according to
\begin{equation}
  \sigma_i(\omega;z_*)
  =\Sigma_i(z_h)+\mathcal O(\epsilon_h),
  \qquad
  \mathcal G_i(\omega;z_*)
  =-i\omega\Sigma_i(z_h)+\mathcal O(\epsilon_h).
  \label{eq:numerical-horizon-initial-data}
\end{equation}
The stability of the result must be verified by decreasing $\epsilon_h$. If higher numerical accuracy is required, the subleading coefficients can be obtained systematically from a Frobenius expansion of Eq.~\eqref{eq:zero-momentum-vector-eom}.

\subsection{Retarded correlator and holographic renormalization}
\label{subsec:spectral-function}

At a finite radial cutoff $z=\epsilon$, the retarded response associated with the cutoff value of the bulk field is
\begin{equation}
  G_{R,i}^{(\epsilon)}(\omega)
  \equiv
  -\frac{\mathcal J_i(\epsilon,\omega)}
  {V_i(\epsilon,\omega)}
  =-i\omega\,\sigma_i(\omega;\epsilon).
  \label{eq:GR-cutoff-sigma}
\end{equation}
For a massive vector, this is not yet the correlator with respect to the canonically normalized boundary source, because the leading asymptotic behaviour differs from that of a Maxwell field.

Assume that the geometry is asymptotically AdS$_5$,
\begin{equation}
  g_{tt}(z),\,g_{x_ix_i}(z),\,g_{zz}(z)
  =\frac{L^2}{z^2}\left[1+o(1)\right],
  \qquad z\to0,
  \label{eq:asymptotic-ads-metric}
\end{equation}
and that the soft-wall profile does not alter the leading indicial equation. The near-boundary expansion of the spatial Proca field is then
\begin{align}
  V_i(z,\omega)
  &=
  z^{1-\nu_X}
  \left[V_i^{(0)}(\omega)+\cdots\right]
  +
  z^{1+\nu_X}
  \left[V_i^{(2\nu_X)}(\omega)+\cdots\right],
  \\
  \nu_X
  &\equiv
  \sqrt{1+m_{5,X}^2L^2}
  =\Delta_X-2.
  \label{eq:massive-vector-UV-nu}
\end{align}
For an integer $\nu_X$, logarithmic terms can appear at the order where the two Frobenius branches overlap. These terms generate local, scheme-dependent contributions and are removed by the standard holographic counterterms.

The source of the dual operator is the coefficient of the non-normalizable mode,
\begin{equation}
  V_i^{(0)}(\omega)
  =
  \lim_{\epsilon\to0}
  \epsilon^{\nu_X-1}V_i(\epsilon,\omega).
  \label{eq:source-cutoff-normalization}
\end{equation}
Consequently, the renormalized retarded correlator is
\begin{equation}
  G_{R,i}^{\mathrm{ren}}(\omega)
  =
  \lim_{\epsilon\to0}
  \left[
  \epsilon^{-2(\nu_X-1)}
  G_{R,i}^{(\epsilon)}(\omega)
  +G_{R,i}^{\mathrm{ct}}(\omega,\epsilon)
  \right],
  \label{eq:GR-ren-source-normalization}
\end{equation}
where $G_{R,i}^{\mathrm{ct}}$ denotes the source-normalized contribution of local boundary counterterms. These counterterms remove ultraviolet divergences and scheme-dependent contact terms. For real nonzero frequency, they do not modify the absorptive part that contains the thermal resonance structure.

We adopt the convention
\begin{equation}
  \rho_{X,i}(\omega)
  \equiv
  -\operatorname{Im}G_{R,i}^{\mathrm{ren}}(\omega).
  \label{eq:spectral-density-def}
\end{equation}
(Some authors include an additional overall factor of $2$ in the definition of the spectral density.) Using Eq.~\eqref{eq:GR-cutoff-sigma}, the spectral function becomes
\begin{equation}
  \rho_{X,i}(\omega)
  =
  \lim_{\epsilon\to0}
  \epsilon^{-2(\nu_X-1)}
  \omega\,\operatorname{Re}\sigma_i(\omega;\epsilon).
  \label{eq:rho-source-normalization}
\end{equation}
Equivalently,
\begin{equation}
  \rho_{X,i}(\omega)
  =
  -\lim_{\epsilon\to0}
  \epsilon^{-2(\nu_X-1)}
  \operatorname{Im}\mathcal G_i(\omega;\epsilon).
  \label{eq:rho-from-response-flow}
\end{equation}
The factor $\epsilon^{-2(\nu_X-1)}$ is fixed by the normalization of the non-normalizable mode and is therefore part of the holographic dictionary; it is not an arbitrary numerical rescaling. At a fixed cutoff and within a single channel, it changes only the overall normalization, but it is required for the continuum limit and for comparisons between operators of different canonical ultraviolet dimension.

In an isotropic background,
\begin{equation}
  G_{R,1}=G_{R,2}=G_{R,3},
  \qquad
  \rho_{X,1}=\rho_{X,2}=\rho_{X,3},
\end{equation}
and we omit the polarization index in what follows. No additional factor of three is included unless an explicitly polarization-summed spectral density is desired.

\subsection{Ultraviolet-rescaled spectral function}
\label{subsec:UV-rescaled-spectral-function}

In four boundary dimensions, the momentum-space two-point function of a spin--$1$ operator of dimension $\Delta_X$ has mass dimension
\begin{equation}
  [G_R]=[\rho_X]=[\mathrm{mass}]^{2\Delta_X-4}.
  \label{eq:spectral-mass-dimension}
\end{equation}
Accordingly, the leading large-frequency behaviour of the absorptive part is
\begin{equation}
  \rho_X(\omega)
  \sim
  \omega^{2\Delta_X-4},
  \qquad
  \omega\to\infty,
  \label{eq:rho-UV-scaling-delta}
\end{equation}
although the renormalized correlator itself may contain logarithms when $\nu_X$ is an integer. To make the resonance structure more visible, we define the dimensionless, UV-rescaled quantity
\begin{equation}
  \widehat\rho_X(\omega)
  \equiv
  \frac{\rho_X(\omega)}{\omega^{2\Delta_X-4}}
  =
  \lim_{\epsilon\to0}
  \frac{
  \epsilon^{-2(\nu_X-1)}
  \omega\,\operatorname{Re}\sigma(\omega;\epsilon)
  }{
  \omega^{2\Delta_X-4}
  }.
  \label{eq:rhohat-source-and-UV-rescaling}
\end{equation}
This quantity is used only for $\omega>0$. The source-normalization factor and the ultraviolet division perform conceptually different operations: the former is required to define the renormalized boundary correlator, whereas the latter is a diagnostic rescaling that removes the dominant high-frequency growth. Thus, although $\widehat{\rho}_X$ is not the physical spectral density itself, it provides a useful diagnostic quantity for characterizing the thermal dissociation of the mesonic state.

For the $\pi_1$ hybrid channel, $\Delta_{\pi_1}=5$ and $\nu_{\pi_1}=3$. Therefore,
\begin{equation}
  \widehat\rho_{\pi_1}(\omega)
  =
  \lim_{\epsilon\to0}
  \frac{
  \epsilon^{-4}
  \omega\,\operatorname{Re}\sigma_{\pi_1}(\omega;\epsilon)
  }{\omega^6}
  =
  \lim_{\epsilon\to0}
  \frac{\epsilon^{-4}\rho_{\pi_1}^{(\epsilon)}(\omega)}
  {\omega^6}.
  \label{eq:rhohat-pi1-source-normalized}
\end{equation}
For a $Z_c$-like tetraquark channel, $\Delta_{Z_c}=6$ and $\nu_{Z_c}=4$, giving
\begin{equation}
  \widehat\rho_{Z_c}(\omega)
  =
  \lim_{\epsilon\to0}
  \frac{
  \epsilon^{-6}
  \omega\,\operatorname{Re}\sigma_{Z_c}(\omega;\epsilon)
  }{\omega^8}
  =
  \lim_{\epsilon\to0}
  \frac{\epsilon^{-6}\rho_{Z_c}^{(\epsilon)}(\omega)}
  {\omega^8}.
  \label{eq:rhohat-zc-source-normalized}
\end{equation}
These are the quantities used below to compare the thermal dissociation of the exotic resonance state after both the holographic source normalization and the leading ultraviolet power law have been taken into account.

\section{Exotic spin--1 channels in the gluon-condensate background}
\label{sec:exotic-gluon-condensate}

In this section, we apply the general membrane-flow formalism derived in section~\ref{sec:spectral-membrane} to two exotic spin--$1$ mesonic channels, namely the hybrid channel $\pi_1$ and the charged charmonium-like channel $Z_c$. The strategy is the following. First, we specify the gluon-condensate background geometry. Second, we fix the channel-dependent soft-wall profile $\Phi_X(z)$, with $X=\pi_1, Z_c$. Third, we solve the zero-temperature Schr\"odinger-like problem in the gluon-condensate background and show how the
vacuum masses acquire a dependence on the gluon-condensate parameter $c$. Finally, we compute the finite-temperature spectral functions.

A central point of the present construction is that the background dilaton $\phi(z)$ and the channel-dependent soft-wall profile $\Phi_X(z)$ play different roles in the holographic description. The field $\phi(z)$ belongs to the Einstein--dilaton background sector and determines the vacuum structure of the boundary theory through its backreaction on the five-dimensional geometry. In particular, the deformation parameter $c$ appearing in $\phi(z)$ is related to the gluon
condensate through the holographic dictionary discussed below. In contrast, $\Phi_X(z)$ is introduced in the probe action of a specific mesonic channel $X$, phenomenologically describes the infrared confinement properties associated with that channel and is therefore fixed once the hadronic state is chosen. It is not a dynamical bulk scalar field and does not represent an additional gluon-condensate degree of freedom. Consequently, $\Phi_X(z)$ is kept independent of $c$. This separation allows
us to isolate the effect of the gluon-condensate deformation on mesonic observables. Although $\Phi_X(z)$ is independent of $c$, the mass spectrum obtained from the Schr\"odinger-like equation remains generally $ c$-dependent because the background metric is modified by the $c$-dependent Einstein--dilaton solution. Only in the limit $c=0$ does the zero-temperature eigenvalue problem reduce to the soft-wall problem used to fix the parameters of $\Phi_X(z)$.

\subsection{Gluon-condensate black-hole geometry}
\label{subsec:gluon-background}

We consider the five-dimensional Einstein-dilaton action
\begin{equation}
S_{\rm grav}
 =
 \frac{1}{2\kappa_5^2}
 \int d^5x \sqrt{-G}
 \left[
 R+\frac{12}{L^2}
 -\frac{1}{2}G^{mn}\partial_m\phi\,\partial_n\phi
 \right],
\label{eq:gluon-gravity-action}
\end{equation}
where $L$ is the AdS radius and $\kappa_5$ is the five-dimensional gravitational coupling. The scalar field $\phi$ is associated with the gluonic operator in the boundary theory. A non-vanishing gluon condensate is represented holographically by a non-trivial background profile of $\phi(z)$.

The dilatonic black-hole solution can be written as~\cite{Kim:2007qk}
\begin{equation}
ds^2
 =
 \frac{L^2}{z^2}
 \left[
 {\cal A}(z)
 \left(
 d\vec{x}^{\,2}
 -
 {\cal H}(z)\,dt^2
 \right)
 + dz^2
 \right],
\label{eq:gluon-condensate-metric-compact}
\end{equation}
where
\begin{align}
{\cal A}(z)
&=
\left(1-f^2z^8\right)^{1/2}
\left(
\frac{1+fz^4}{1-fz^4}
\right)^{\frac{a}{2f}},
\label{eq:Acal-def}
\\
{\cal H}(z)
&=
\left(
\frac{1-fz^4}{1+fz^4}
\right)^{\frac{2a}{f}} .
\label{eq:Hcal-def}
\end{align}
The parameters $f$, $a$ and $c$ are related by
\begin{equation}
f^2=a^2+c^2,
\qquad
a=\frac{1}{4}(\pi T)^4 .
\label{eq:f-a-c-relation}
\end{equation}
Here $c$ denotes the gluon-condensate parameter, whose quantitative relation to the gluon-condensate expectation value is established below through the near-boundary holographic dictionary. The geometry is defined in the radial interval
\begin{equation}
0<z<z_f,
\qquad
z_f=f^{-1/4}.
\label{eq:zf-def}
\end{equation}
In the notation of the general diagonal metric used in section~\ref{sec:spectral-membrane}, the positive metric functions are
\begin{equation}
g_{tt}(z)
 =
 \frac{L^2}{z^2}{\cal A}(z){\cal H}(z),
\qquad
g_{xx}(z)
 =
 \frac{L^2}{z^2}{\cal A}(z),
\qquad
g_{zz}(z)
 =
 \frac{L^2}{z^2}.
\label{eq:metric-components-gluon}
\end{equation}
The background dilaton is
\begin{equation}
\phi(z)
 =
 \frac{c}{f}
 \sqrt{\frac{3}{2}}\,
 \log
 \left(
 \frac{1+fz^4}{1-fz^4}
 \right)
 +\phi_0 ,
\label{eq:background-dilaton}
\end{equation}
where $\phi_0$ is an integration constant. Near the asymptotic boundary,
\begin{equation}
  \phi(z)
  =
  \phi_0
  +
  \sqrt{6}\,c\,z^4
  +
  \mathcal O(z^8).
\end{equation}
According to the holographic dictionary~\cite{Csaki:2006ji,Ko:2009jc}, the coefficient of the normalizable $z^4$ mode is proportional to the expectation value of the dimension-four gluonic operator. The gluon-condensate parameter $c$ is quantitatively related to the conventional dimension-four gluon condensate by
\begin{equation}\label{eq:gluon_condensate_c}
\left\langle\frac{\alpha_s}{\pi}G^2\right\rangle=\frac{8\alpha_s\sqrt{3(N_c^2-1)}}{\pi^2}\,c ,
\end{equation}
where $\alpha_s$ is the gauge coupling. For $N_c=3$, Eq.~(\ref{eq:gluon_condensate_c}) reduces to
\begin{equation}\label{eq:gluon_condensate_c_Nc3}
\left\langle\frac{\alpha_s}{\pi}G^2\right\rangle=\frac{16\sqrt{6}\,\alpha_s}{\pi^2}\,c .
\end{equation}

It is important to distinguish the background field $\phi(z)$ from the soft-wall profile $\Phi_X(z)$ appearing in the probe-vector action. The former is a dynamical field in the Einstein--dilaton gravitational sector and encodes the gluonic vacuum deformation through its normalizable mode. The latter is a phenomenological function introduced in the probe action to characterize the infrared confinement properties of a specific mesonic channel. Therefore, these two functions are associated with different physical effects and should not be identified. The parameter $c$ therefore affects the mesonic observables through the metric functions, not through a refitting of $\Phi_X(z)$.

At zero temperature, one has $a=0$ and $f=c$. The geometry reduces to the dilaton wall solution,
\begin{equation}
ds_0^2
 =
 \frac{L^2}{z^2}
 \left[
 W(z)
 \left(
 -dt^2+d\vec{x}^{\,2}
 \right)
 +dz^2
 \right],
\qquad
W(z)=\sqrt{1-c^2z^8}.
\label{eq:zero-temperature-gluon-wall}
\end{equation}
The zero-temperature geometry is defined on
\begin{equation}
0<z<z_c,
\qquad
z_c=c^{-1/4}.
\label{eq:zc-def}
\end{equation}
In the limit $c\to0$, one has $W(z)\to1$ and $z_c\to\infty$, so that the background reduces to pure AdS.

In the present Einstein--dilaton model, the confined and deconfined phases are represented by the thermal dilatonic geometry and the dilatonic black-hole geometry, respectively. The difference between their renormalized Euclidean on-shell action densities is given by~\cite{Zhu:2021vkj}
\begin{equation}
\Delta F
\equiv
F_{\mathrm{dBH}}-F_{\mathrm{tdAdS}}
=
\frac{L^{3}}{2\kappa_{5}^{2}}
\left(3a+4c-4f\right).
\label{eq:free_energy_difference}
\end{equation}
The Hawking--Page transition is determined by $\Delta F(T_{c})=0$. Apart from the trivial zero-temperature solution, the transition condition gives
\begin{equation}
a_{c}=\frac{24}{7}c,
\qquad
T_{c}(c)
=
\frac{1}{\pi}
\left(\frac{96c}{7}\right)^{1/4}.
\label{eq:critical_temperature}
\end{equation}
The phenomenological value of the dimension-four gluon condensate is commonly taken to be $\left\langle\frac{\alpha_s}{\pi}G^2\right\rangle\simeq 0.012~{\rm GeV}^4$ from QCD sum-rule analyses \cite{Shifman:1978bx}. Using $\alpha_s=0.73$, as adopted in \cite{Zhu:2021vkj}, Eq.~\eqref{eq:gluon_condensate_c_Nc3} gives $c\simeq 0.004~{\rm GeV}^4$. We therefore take $c=0.004~{\rm GeV}^4$ as the phenomenologically motivated upper benchmark in the present calculation. This value corresponds to
$\left\langle\frac{\alpha_s}{\pi}G^2\right\rangle\simeq 0.0116~{\rm GeV}^4$,
which is in good agreement with the QCD sum-rule value.

As an independent thermodynamic consistency check, the same value $c=0.004~{\rm GeV}^4$ gives $T_c \simeq 154.0~{\rm MeV}$, which is consistent with the deconfinement temperature $154\pm9~\mathrm{MeV}$ obtained from lattice simulations \cite{Bazavov:2011nk,HotQCD:2014kol}, the chemical freeze-out temperature
$156.5\pm1.5~\mathrm{MeV}$ extracted from relativistic heavy-ion collisions
\cite{Andronic:2017pug}, and the value $155.38~\mathrm{MeV}$ obtained previously using the phenomenological gluon-condensate input from QCD sum rules~\cite{Zhu:2021vkj}. On the other hand, the benchmark range of $c$ should also be selected by vacuum-spectral considerations.

\subsection{QCD interpolating currents and Fixed channel profiles for \texorpdfstring{$\pi_1$}{pi1} and \texorpdfstring{$Z_c$}{Zc}}
\label{subsec:fixed-channel-profiles}

To specify the ultraviolet quantum numbers of the two exotic spin--1 channels, we introduce representative local QCD interpolating currents. For the isovector hybrid channel, we adopt the dimension-five current
\begin{equation}
J_{\mu}^{a,\pi_1}(x)
=
\bar q(x) T^{a} G_{\mu\nu}(x)\gamma^{\nu}q(x),
\label{eq:pi1_current}
\end{equation}
where $q=(u,d)^{T}$ denotes the light-quark doublet, $T^{a}$ is a flavor generator, and $G_{\mu\nu}=G_{\mu\nu}^{A}t^{A}$ is the gluon field-strength tensor in color space. This current carries the exotic quantum numbers $J^{PC}=1^{-+}$. Its canonical ultraviolet dimension is
\begin{equation}
\Delta_{\pi_1}
=
[\bar q]+[G_{\mu\nu}]+[q]
=
\frac{3}{2}+2+\frac{3}{2}
=
5.
\end{equation}
It is therefore dual to a massive one-form field satisfying $m_{5,\pi_1}^{2}L^{2}=(\Delta_{\pi_1}-1)(\Delta_{\pi_1}-3)=8$ \cite{Bellantuono:2014lra}.

For the charged charmonium-like $Z_c$ channel, a representative local axial-vector four-quark current can be written as
\begin{equation}
J_{\mu}^{Z_c^+}(x)=\frac{\epsilon^{ijk}\epsilon^{imn}}{\sqrt{2}}
\Big[u_{j}^{T}(x)C\gamma_{5}c_{k}(x)\,\bar d_{m}(x)\gamma_{\mu}C\bar c_{n}^{T}(x)-u_{j}^{T}(x)C\gamma_{\mu}c_{k}(x)\,\bar d_{m}(x)\gamma_{5}C\bar c_{n}^{T}(x)\Big],
\label{eq:zc_current}
\end{equation}
where $i,j,k,m,n$ are color indices and $C$ is the charge-conjugation matrix. The charged current has $J^{P}=1^{+}$, while its neutral isospin partner has $C=-$; the corresponding channel is conventionally denoted by $I^{G}(J^{PC})=1^{+}(1^{+-})$. Since the current contains four quark fields, its canonical ultraviolet dimension is
\begin{equation}
\Delta_{Z_c}
=
4\times\frac{3}{2}
=
6,
\end{equation}
which gives $m_{5,Z_c}^{2}L^{2}=(\Delta_{Z_c}-1)(\Delta_{Z_c}-3)=15$ \cite{Wang:2013vex}.

The local four-quark current is not unique, since currents with diquark--antidiquark and meson--meson color structures can couple to the same set of quantum numbers. In the present bottom-up construction, Eqs.~\eqref{eq:pi1_current} and \eqref{eq:zc_current} should therefore be regarded as representative operators used to fix the ultraviolet quantum numbers and canonical dimensions of the corresponding holographic channels. The model does not by itself uniquely determine the microscopic internal structure of the $Z_c$ state. Possible anomalous dimensions and mixing among operators with identical quantum numbers are neglected.

Having specified the ultraviolet operator assignments, we now introduce the channel-dependent profiles that control the infrared spectral properties, which is primarily responsible for the properties of different mesons. The profile entering the probe action~\eqref{eq:proca-action} is chosen in the non-quadratic form~\cite{MartinContreras:2021bis}:
\begin{equation}
\Phi_X(z)
 =
(\kappa_X z)^{2-\alpha_X}
+M_X z
+\tanh
\left(
\frac{1}{M_Xz}
-
\frac{\kappa_X}{\sqrt{\Gamma_X}}
\right).
\label{eq:PhiX-profile}
\end{equation}
The parameters
\begin{equation}
\Theta_X
 =
\{\kappa_X,M_X,\sqrt{\Gamma_X},\alpha_X\}
\label{eq:ThetaX-def}
\end{equation}
are not fitted again as functions of $c$. Instead, for a fixed mesonic channel $X$, we use the values obtained in the zero-condensate soft-wall calibration. Here, the channel-dependent profile $\Phi_X(z)$ is regarded as an effective probe-sector parametrization within the bottom-up holographic framework, rather than as an additional dynamical background field. Thus, $\partial_c \Phi_X(z)=0$, the gluon-condensate dependence of the mass spectrum and spectral function will arise entirely from the background metric \footnote{In current holographic probe constructions, we take the minimal probe coupling, in which background fields affect mesonic observables through the deformed bulk geometry. Of course, non-minimal interactions may include an additional explicit coupling between the background field and the probe sector, but this is not required for the consistency of the present probe construction.}.

The channel assignments, canonical ultraviolet dimensions, bulk Proca masses, and fixed soft-wall parameters are summarized in table~\ref{tab:channel-parameters}. The soft-wall parameters are taken from the zero-condensate calibration of~\cite{Toniato:2025gts}, whereas the values of $\Delta_X$ follow from the representative QCD interpolating currents specified above.  With these parameters, the $c=0$ zero-temperature spectrum reproduces the corresponding vacuum soft-wall masses. For $c\neq0$, the same profiles $\Phi_{\pi_1}(z)$ and $\Phi_{Z_c}(z)$ are used, but the resulting eigenvalues shift because the Schr\"odinger potential is modified by the gluon-condensate geometry. With the input listed in table~\ref{tab:channel-parameters}, the only remaining parameter controlling the zero-temperature Einstein--dilaton background is the gluon-condensate parameter $c$. The dependence of the vacuum masses on $c$ is studied in the next subsection.

\begin{table}[t]
\centering
\begin{tabular}{c c c c c c}
\hline
Channel & Interpretation & $\Delta_X$ & $m_{5,X}^2L^2$
& $\nu_X$ & Parameters fixed from $c=0$ spectrum \\
\hline
$\pi_1$ & hybrid meson & $5$ & $8$ & $3$
& $\kappa=0.468$, $M=0.20$, $\sqrt{\Gamma}=0.12$, $\alpha=0.034$ \\
$Z_c$ & tetraquark-like state & $6$ & $15$ & $4$
& $\kappa=1.75$, $M=1.44$, $\sqrt{\Gamma}=0.30$, $\alpha=0.539$ \\
\hline
\end{tabular}
\caption{Channel assignments, canonical ultraviolet dimensions, bulk Proca masses, and fixed soft-wall parameters. The dimensions $\Delta_X$ correspond to the representative
local QCD interpolating currents in Eqs.~\eqref{eq:pi1_current} and \eqref{eq:zc_current}. The parameters $\kappa_X$, $M_X$, and $\sqrt{\Gamma_X}$ are given in GeV, whereas $\alpha_X$, $\Delta_X$, $m_{5,X}^{2}L^{2}$, and $\nu_X$ are dimensionless.}
\label{tab:channel-parameters}
\end{table}

\subsection{Zero-temperature mass spectrum and its \texorpdfstring{$c$}{c} dependence}
\label{subsec:zero-temperature-spectrum-c}

We first determine the zero-temperature spectrum in the background \eqref{eq:zero-temperature-gluon-wall}. We work in the transverse channel and use the normal-mode ansatz
\begin{equation}
V_\mu(x,z)
 =
 \epsilon_\mu v_n(z)e^{ik\cdot x},
\qquad
k^2=-m_n^2,
\qquad
k^\mu\epsilon_\mu=0.
\label{eq:normal-mode-ansatz-c}
\end{equation}
For a transverse spatial component, the equation of motion following from \eqref{eq:proca-action} can be written as the Sturm--Liouville problem
\begin{equation}
\partial_z
\left[
P_X(z;c)\partial_z v_n(z)
\right]
+
\left[
m_n^2 Q_X(z)-R_X(z;c)
\right]v_n(z)=0,
\label{eq:SL-c}
\end{equation}
with
\begin{align}
P_X(z;c)
&=
\sqrt{-G}\,e^{-\Phi_X(z)}G^{zz}G^{ii}
 =
\frac{L}{z}e^{-\Phi_X(z)}W(z),
\label{eq:P-c}
\\
Q_X(z)
&=
\sqrt{-G}\,e^{-\Phi_X(z)}|G^{tt}|G^{ii}
 =
\frac{L}{z}e^{-\Phi_X(z)},
\label{eq:Q-c}
\\
R_X(z;c)
&=
\sqrt{-G}\,e^{-\Phi_X(z)}m_{5,X}^2G^{ii}
 =
\frac{L^3}{z^3}e^{-\Phi_X(z)}W(z)m_{5,X}^2 .
\label{eq:R-c}
\end{align}
This equation shows explicitly that the eigenvalues depend on $c$ even though $\Phi_X(z)$ itself does not.

To cast \eqref{eq:SL-c} into Schr\"odinger form, we introduce the Liouville coordinate
\begin{equation}
\frac{dr}{dz}
 =
\sqrt{\frac{Q_X(z)}{P_X(z;c)}}
 =
\frac{1}{\sqrt{W(z)}}
 =
\frac{1}{(1-c^2z^8)^{1/4}},
\label{eq:r-z-c}
\end{equation}
or
\begin{equation}
r(z;c)
 =
\int_0^z
\frac{d\xi}{(1-c^2\xi^8)^{1/4}}.
\label{eq:r-integral-c}
\end{equation}
The IR endpoint is mapped to
\begin{equation}
r_{\rm IR}(c)
 =
r(z_c;c)
 =
\int_0^{c^{-1/4}}
\frac{d\xi}{(1-c^2\xi^8)^{1/4}}.
\label{eq:rIR-c}
\end{equation}
In the limit $c\to0$, one obtains $r(z;c)\to z$ and $r_{\rm IR}\to\infty$.

In terms of $r$, the Sturm--Liouville equation becomes
\begin{equation}
\partial_r
\left[
C_X(r;c)\partial_r v_n(r)
\right]
+
C_X(r;c)
\left[
m_{X,n}^2-M_{5,X}^2(r;c)
\right]v_n(r)=0,
\label{eq:SL-r-c}
\end{equation}
where
\begin{equation}
C_X(r;c)
 =
\sqrt{P_X(z;c)Q_X(z)}
 =
\frac{L}{z(r;c)}
e^{-\Phi_X(z(r;c))}
\sqrt{W(z(r;c))}
\label{eq:CX-c}
\end{equation}
and
\begin{equation}
M_{5,X}^2(r;c)
 =
\frac{R_X(z;c)}{Q_X(z)}
 =
m_{5,X}^2L^2
\frac{W(z(r;c))}{z^2(r;c)} .
\label{eq:M5-c}
\end{equation}
Although the expression $\Phi_X(z(r;c))$ contains $c$ through the coordinate map $z=z(r;c)$, this is not a refitting or intrinsic medium dependence of $\Phi_X$. The function $\Phi_X(z)$ is fixed once and for all in the original holographic coordinate $z$.

Defining
\begin{equation}
v_n(r)=e^{-H_X(r;c)}\psi_n(r),
\qquad
H_X(r;c)=\frac{1}{2}\log C_X(r;c),
\label{eq:H-c}
\end{equation}
one obtains
\begin{equation}
-\psi_n''(r)+U_X(r;c)\psi_n(r)
 =
m_{X,n}^2(c)\psi_n(r),
\label{eq:Schrodinger-c}
\end{equation}
with
\begin{equation}
U_X(r;c)
 =
\left[H_X'(r;c)\right]^2
+
H_X''(r;c)
+
m_{5,X}^2L^2
\frac{W(z(r;c))}{z^2(r;c)}.
\label{eq:potential-c}
\end{equation}
The normalizable modes satisfy
\begin{equation}
\psi_n(r\to0)\sim r^{\nu_X+\frac{1}{2}},
\qquad
\psi_n(r_{\rm IR})=0 .
\label{eq:BC-c}
\end{equation}
In practice, the second condition is imposed at a numerical endpoint close to $r_{\rm IR}(c)$.

The vacuum masses in the gluon-condensate background are therefore
\begin{equation}
M_{X,n}^{\rm vac}(c)=m_{X,n}(c).
\label{eq:vacuum-masses-c}
\end{equation}

The numerical results for the zero-temperature masses are presented in table~\ref{tab:vacuum-masses-c}. The column at $c=0$ serves as a consistency check, since it reproduces the reference soft-wall spectrum obtained with the same fixed channel-dependent profile $\Phi_X(z)$~\cite{Toniato:2025gts}. The columns with finite $c$ then show how the background geometry associated with the gluon condensate modifies the vacuum spectrum when the profile parameters are kept unchanged.

\begin{table}[t]
\centering
\setlength{\tabcolsep}{6pt}
\renewcommand{\arraystretch}{1.15}

\begin{tabular}{llcccccc}
\hline
\multirow{2}{*}{Channel}
& \multirow{2}{*}{State}
& \multicolumn{1}{c|}{\multirow{2}{*}{$M^{\rm exp}$[${\rm MeV}$]}}
& \multicolumn{5}{c}{$M^{\rm vac}(c)$[${\rm MeV}$]} \\
\cline{4-8}
& &
\multicolumn{1}{c|}{}
& $c=0$
& $c=0.001$
& $c=0.002$
& $c=0.003$
& $c=0.004$ \\
\hline

\multirow{3}{*}{$\pi_1$}
& $\pi_1(1400)$ & $1354\pm25$  & $1354.7$ & $1393.7$ & $1469.7$ & $1535.4$ & $1591.2$ \\
& $\pi_1(1600)$ & $1660^{+15}_{-11}$ &$1617.8$ & $1793.5$ & $1989.7$ & $2146.4$ & $2280.3$ \\
& $\pi_1(2015)$ & $2014\pm20\pm16$ & $1847.2$ & $2271.5$ & $2631.2$ & $2905.3$ & $3131.2$ \\

\hline

\multirow{3}{*}{$Z_c$}
& $Z_c(3900)$ & $3887.2\pm2.3$ & $3890.3$ & $3889.6$ & $3889.3$ & $3892.3$ & $3899.0$ \\
& $Z_c(4200)$ & $4196^{+35}_{-32}$ & $4190.3$ & $4188.9$ & $4209.3$ & $4252.2$ & $4303.6$ \\
& $Z_c(4430)$ & $4478^{+15}_{-18}$ & $4446.6$ & $4462.0$ & $4567.0$ & $4689.3$ & $4806.4$ \\

\hline
\end{tabular}
\caption{
Zero-temperature masses of the $\pi_1$ and $Z_c$ channels. The values under $M^{\rm vac}(c)$ correspond to different gluon-condensate parameters $c$ in ${\rm GeV}^4$. Experimental results $M^{\rm exp}$ are from PDG~\cite{ParticleDataGroup:2018ovx}.
}
\label{tab:vacuum-masses-c}
\end{table}

The phenomenological gluon-condensate matching discussed section \ref{subsec:gluon-background} motivates a upper bound, $c=0.004~\mathrm{GeV}^{4}$. We further require that the same background deformation does not produce an unacceptably large modification of the vacuum ground-state spectrum before temperature is introduced. To quantify this condition, we define the relative ground-state mass shift as
\begin{equation}
\delta_X^{(0)}(c) = \left| \frac{ M_{X,0}^{\rm vac}(c)-M_{X,0}^{\rm vac}(0) } {M_{X,0}^{\rm vac}(0)} \right|<\delta_{\rm max},
\label{eq:ground-shift-c}
\end{equation}
where $X=\pi_1,Z_c$. In the present work, we take $\delta_{\max}=20\%$ as a practical tolerance for the ground-state vacuum-mass deformation. This tolerance is not intended as a universal physical bound, but as an internal consistency criterion ensuring that the channel calibrated at $c=0$ remains continuously identifiable over the selected range of $c$.

At the upper benchmark $c=0.004~\mathrm{GeV}^{4}$, the ground-state mass shift is approximately $17.5\%$ for the $\pi_1$ channel, whereas it remains below $0.3\%$ for the $Z_c$ channel. Thus, the complete range $c\leq0.004~\mathrm{GeV}^{4}$ satisfies the adopted ground-state mass-tolerance criterion. Together with the fact that $c=0.004~\mathrm{GeV}^{4}$ yields a transition temperature close to the lattice and heavy-ion phenomenological scales, this provides a combined thermodynamic and spectroscopic justification for the benchmark interval used in the subsequent finite-temperature analysis.

From table~\ref{tab:vacuum-masses-c}, one can see that the influence of $c$ is channel dependent. In the $\pi_1$ channel, the masses increase monotonically with $c$, and the excited states are more sensitive to the gluon-condensate deformation than the ground state. For the ground-state $\pi_1(1400)$, the relative shifts are about $2.9\%$, $8.5\%$, $13.3\%$, and $17.5\%$ for $c=0.001$, $0.002$, $0.003$, and $0.004~{\rm GeV}^4$, respectively. The larger deviations of the excited states indicate that higher modes are more sensitive to the background deformation. The $Z_c$ channel shows a much weaker dependence on $c$. The ground-state $Z_c(3900)$ is almost unchanged in the whole range of $c$ considered here, with relative shifts below $0.3\%$. The excited states receive somewhat larger corrections than the ground state, but the overall distortion is still much smaller than that in the $\pi_1$ channel. This suggests that the $Z_c$ channel is more stable under the gluon-condensate deformation. Therefore, although some excited states, especially in the $\pi_1$ channel, exhibit sizable mass shifts at larger $c$, in this work, our main interest is the thermal behavior of the ground-state mesons.

\subsection{Finite-temperature spectral functions}
\label{subsec:finite-temperature-spectrum}

For a fixed benchmark value $c=c_0$, the thermal spectral function $\widehat{\rho}_X(\omega;T,c_0)$ describes the thermal evolution of the same vacuum channel whose zero-temperature mass is $M_{X,n}^{\rm vac}(c_0)$. In particular, in the low-temperature limit, the resonance peak is expected to approach the vacuum mass $M_{X,n}^{\rm vac}(c_0)$ rather than the undeformed $c=0$ spectrum.

At zero temperature, the spectrum is obtained from the normal-mode quantization of the bulk field. Once the temperature is introduced, the background contains a black-hole horizon described by~\eqref{eq:gluon-condensate-metric-compact}, and the discrete normal-mode problem is replaced by the computation of the retarded correlator with incoming-wave boundary conditions at the horizon. Therefore, the position of resonance peaks extracted from the real-frequency spectral function are related to the thermal masses. The thermal width is estimated from the full width at half maximum (FWHM). To provide a quantitative definition of dissociation, we use the temperature dependence of the ground-state peak height in the spectral function. For each channel $X=\pi_1,Z_c$, the ground-state peak is continuously tracked from the low-temperature spectrum, and its height is defined as
\begin{equation}
H_X(T,c)
=
\widehat{\rho}_X
\!\left(
\omega_{p,X}(T,c);T,c
\right),
\end{equation}
where $\omega_{p,X}(T,c)$ denotes the position of the tracked ground-state maximum. Using the lowest temperature in the numerical calculation, $T_0=1~\mathrm{MeV}$, as the reference point, we define the normalized peak height (the same criterion has already been adopted in ~\cite{Bellantuono:2014lra})
\begin{equation}
R_X(T,c)
=
\frac{H_X(T,c)}
     {H_X(T_0,c)} .
\end{equation}
The dissociation temperature is defined as the first temperature at which
\begin{equation}
5\% \leq R_X\bigl(T_{\rm diss}^{X},c\bigr)
\leq 6.67\% ,
\label{eq:dissociation-height-criterion}
\end{equation}
where $T_{\rm diss}^{X}$ denotes dissociation temperature. Since the UV-rescaled spectral functions used here approach an approximately vanishing baseline away from the resonance region, no additional background subtraction is introduced.

For the isotropic background \eqref{eq:metric-components-gluon}, the local membrane factor~\eqref{eq:sigma-i-def} for the channel $X$ is
\begin{equation}
\Sigma_X(z)
 =
\frac{1}{g_5^2}
e^{-\Phi_X(z)}
\sqrt{
\frac{-G}{g_{zz}(z)g_{tt}(z)}
}
g^{xx}(z)
 =
\frac{L}{g_5^2z}
e^{-\Phi_X(z)}
\sqrt{{\cal A}(z)} .
\label{eq:Sigma-X-gluon}
\end{equation}
At zero spatial momentum, the transverse and longitudinal conductivities coincide. The membrane-flow equation becomes
\begin{equation}
\partial_z\sigma_X(\omega;z)
 =
 -i\omega
 \sqrt{\frac{g_{zz}(z)}{g_{tt}(z)}}
 \left[
 \frac{\sigma_X^2(\omega;z)}{\Sigma_X(z)}
 -
 \Sigma_X(z)
 \left(
 1-\frac{m_{5,X}^2g_{tt}(z)}{\omega^2}
 \right)
 \right].
\label{eq:flow-X-finiteT}
\end{equation}
The incoming-wave condition fixes
\begin{equation}
\sigma_X(\omega;z_f)=\Sigma_X(z_f).
\label{eq:horizon-BC-X}
\end{equation}
Numerically, the integration is started at $z=z_f-\epsilon_h$ and evolved to a UV cutoff $z=\epsilon$. Unless otherwise stated, we set $L=1$ and $g_5=1$. The latter only fixes the overall normalization of the spectral function and does not affect the peak positions or widths.

\subsubsection{\texorpdfstring{$\pi_1$}{pi1} channel}
\label{num-pi}

\begin{figure}[t]
\centering
\includegraphics[width=0.6\textwidth]{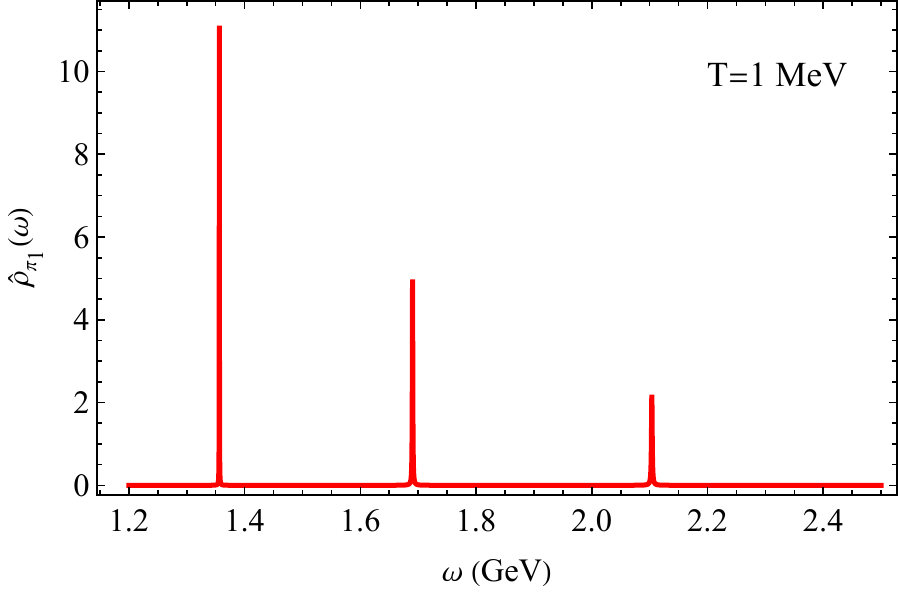}
\caption{
Spectral function $\hat{\rho}_{\pi_1}(\omega)$ of the hybrid
$\pi_1$ channel at $T=1~{\rm MeV}$ and $c=0.001~{\rm GeV}^4$.
}
\label{fig:pi1-lowT}
\end{figure}

We first discuss the thermal spectral function of the hybrid $\pi_1$ channel. Figure~\ref{fig:pi1-lowT} displays the $\pi_1$ spectral function at $T=1~{\rm MeV}$  \footnote{Below the Hawking--Page transition temperature, the black-hole solution provides a metastable continuation of the deconfined thermal branch and remains a well-defined dissipative environment for calculating
retarded correlators with infalling boundary conditions. Consequently, the value of \(T^X_{\mathrm{diss}}\) obtained subsequently marks the temperature at which the vector-meson resonance peak vanishes along the thermal evolution in the black-hole background, thereby characterizing the thermal stability limit of the resonance in the holographic medium considered.} and $c=0.001~{\rm GeV}^4$. At such a low temperature, the thermal spectrum remains close to the corresponding vacuum spectrum. The resonance peak at the smallest frequency can be identified with the ground-state $\pi_1(1400)$ in table~\ref{tab:vacuum-masses-c}, whereas the second and third peaks are associated with the excited states $\pi_1(1600)$ and $\pi_1(2015)$, respectively. The peaks are still narrow and clearly separated, indicating that the thermal width is very small in this near-vacuum regime.

\begin{figure}[t]
\centering
\includegraphics[width=0.48\textwidth]{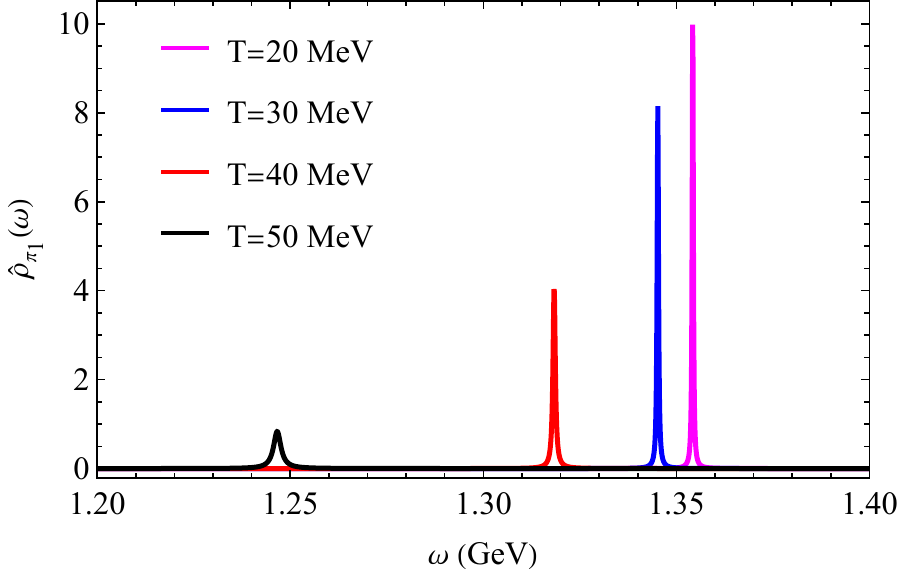}
\includegraphics[width=0.48\textwidth]{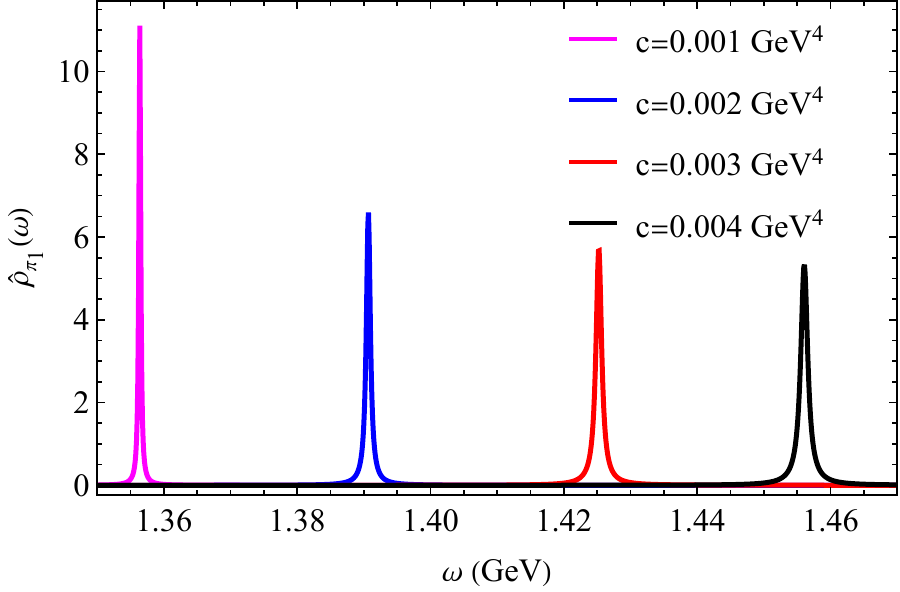}
\caption{Spectral function $\hat{\rho}_{\pi_1}(\omega)$ of the hybrid $\pi_1$ channel. The left panel illustrates the temperature evolution at fixed gluon-condensate parameter $c=0.001~{\rm GeV}^4$, while the right panel shows the modification induced by varying $c$ at fixed temperature $T=1~{\rm MeV}$.}
\label{fig:pi1}
\end{figure}

The temperature dependence of the ground-state peak at fixed $c=0.001~{\rm GeV}^4$ is shown in the left panel of figure~\ref{fig:pi1}. As the temperature increases from $20~{\rm MeV}$ to $50~{\rm MeV}$, the peak moves to lower frequency and becomes broader. This behaviour indicates that temperature reduces the thermal mass and increases the thermal width of the $\pi_1$ ground state. In addition, the peak height decreases significantly with increasing temperature.

The right panel of figure~\ref{fig:pi1} displays the dependence of the $\pi_1$ ground-state peak on the gluon-condensate parameter at fixed $T=1~{\rm MeV}$. The peak position increases as $c$ is enlarged, which is consistent with the zero-temperature mass shift shown in table~\ref{tab:vacuum-masses-c}. This confirms that, at low temperature, the thermal spectral peak is continuously connected to the corresponding $c$-dependent vacuum state. The peak remains well defined for all values of $c$ considered here.

Overall, the $\pi_1$ channel is rather sensitive to both temperature and the gluon-condensate deformation. Increasing $T$ shifts the ground-state peak to a lower frequency, reduces the peak height, and increases the thermal width. These features indicate the gradual loss of the ground-state resonance structure with increasing temperature. A completely similar behavior of hybrid mesons has also been reported in holographic studies of exotic $1^{-+}$ states~\cite{Bellantuono:2014lra}. On the other hand, increasing $c$ shifts the low-temperature ground-state peak to higher frequency, in agreement with the upward shift of the vacuum mass shown in table~\ref{tab:vacuum-masses-c}. By continuously tracking the ground-state peak and applying the normalized-height criterion~\eqref{eq:dissociation-height-criterion}, we find that \[T_{\rm diss}^{\pi_1}=51,\ 65,\ 75,\ 83~{\rm MeV}\] for \[c=0.001,\ 0.002,\ 0.003,\ 0.004~{\rm GeV}^4,\] respectively. The relatively fragile thermal behavior of the hybrid $\pi_1$ state is qualitatively consistent with previous holographic analyses showing earlier melting of hybrid mesons compared with conventional mesons~\cite{Bellantuono:2014lra}. These values are listed in table~\ref{tab:dissociation-temperatures}. Thus, within the present criterion, the gluon-condensate parameter delays the disappearance of the $\pi_1$ ground-state peak. This behavior should be understood as the result of two competing effects: the gluon-condensate deformation modifies the spectral shape, while it also raises the corresponding vacuum ground-state mass scale. Therefore, a larger value of $c$ does not necessarily imply a lower absolute dissociation temperature.

\subsubsection{\texorpdfstring{$Z_c$}{Zc} channel}
\label{num-Zc}

\begin{figure}[t]
\centering
\includegraphics[width=0.6\textwidth]{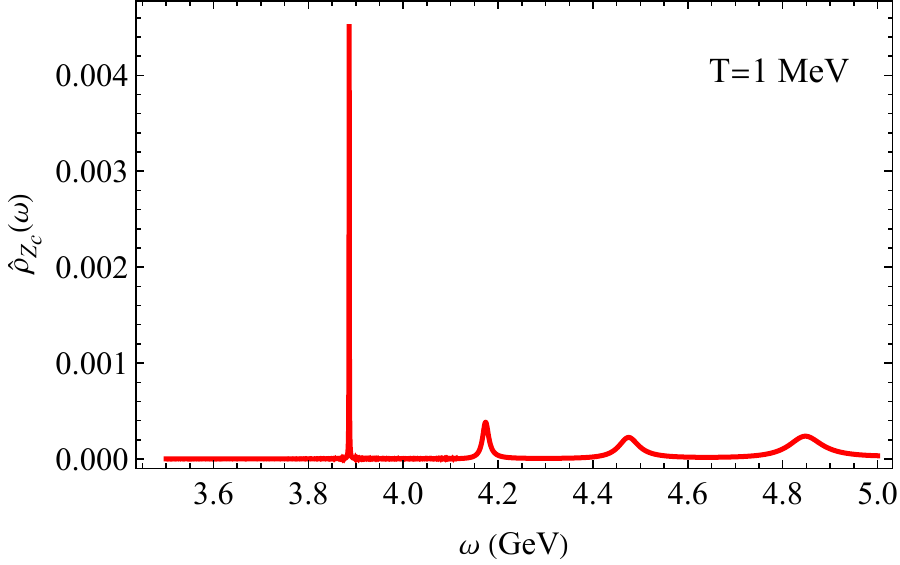}
\caption{Spectral function $\hat{\rho}_{Z_c}(\omega)$ of the tetraquark-like $Z_c$ channel at $T=1~{\rm MeV}$ and $c=0.001~{\rm GeV}^4$.}
\label{fig:zc-lowT}
\end{figure}

We now turn to the charged charmonium-like $Z_c$ channel. Figure~\ref{fig:zc-lowT} shows the spectral function at $T=1~{\rm MeV}$ and $c=0.001~{\rm GeV}^4$. A pronounced narrow peak appears near the ground-state $Z_c(3900)$ mass, while higher structures are also visible at larger frequencies. The ground-state peak is much sharper than the higher-frequency structures, which makes it possible to track its thermal evolution reliably.

\begin{figure}[t]
\centering
\includegraphics[width=0.48\textwidth]{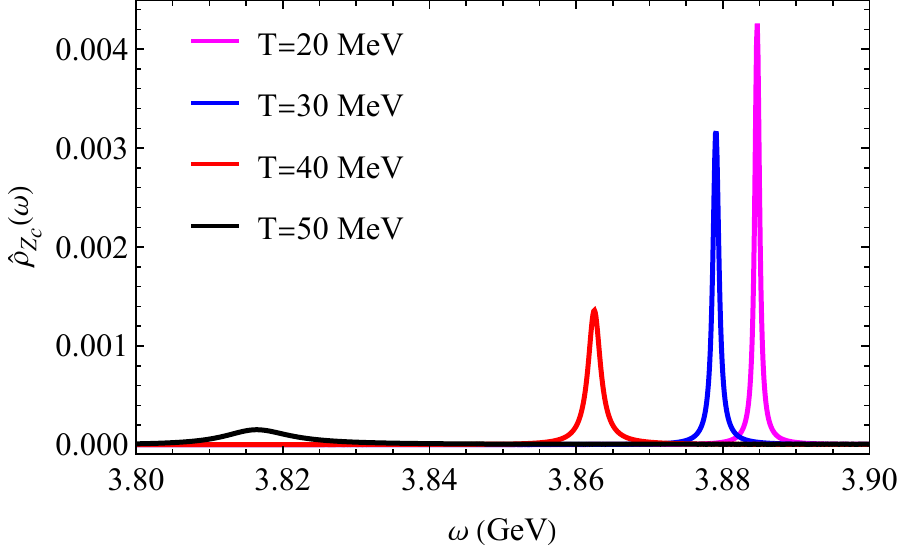}
\includegraphics[width=0.48\textwidth]{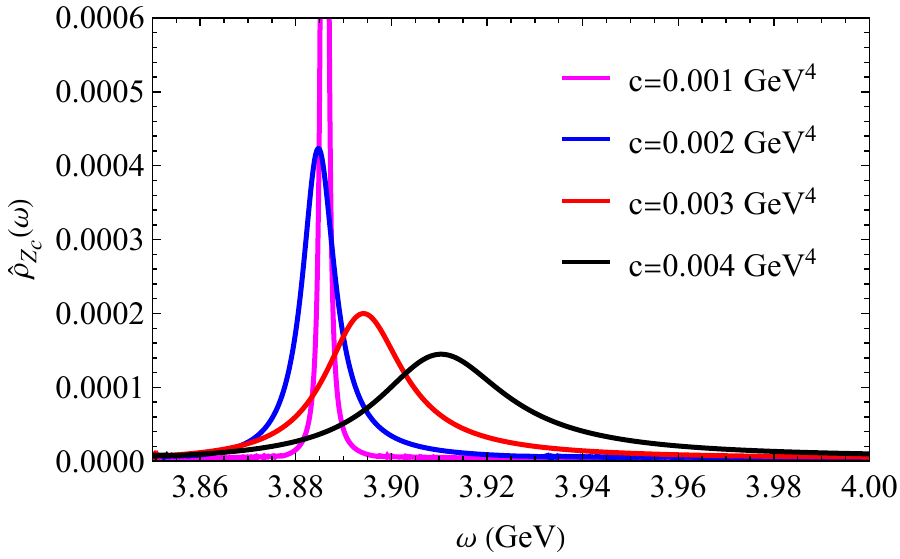}
\caption{Spectral function $\hat{\rho}_{Z_c}(\omega)$ for the tetraquark-like $Z_c$ channel. The left panel illustrates the temperature evolution at fixed gluon-condensate parameter $c=0.001~{\rm GeV}^4$, while the right panel shows the modification induced by varying $c$ at fixed temperature $T=1~{\rm MeV}$.}
\label{fig:zc}
\end{figure}

The temperature dependence at fixed $c=0.001~{\rm GeV}^4$ is shown in the left panel of figure~\ref{fig:zc}. As the temperature increases, the ground-state peak shifts slightly to a lower frequency and its width increases. At the same time, the peak height is reduced. These changes indicate a dissociation of the resonance state and an increase of the thermal width in the medium. The right panel of figure~\ref{fig:zc} shows the dependence of the $Z_c$ ground-state spectral function on the gluon-condensate parameter at fixed $T=1~{\rm MeV}$. The peak position changes only mildly when $c$ is varied from $0.001$ to $0.004~{\rm GeV}^4$, in agreement with the small zero-temperature ground-state mass shifts reported in table~\ref{tab:vacuum-masses-c}. The main effect of increasing $c$ is a visible modification of the spectral shape, including a change in the peak width and height. However, this low-temperature spectral deformation should not be directly interpreted as a lower dissociation temperature. The dissociation temperature should be determined by using the operational criterion; we find \[T_{\rm diss}^{Z_c}=48,\ 61,\ 72,\ 82~{\rm MeV}\] for \[c=0.001,\ 0.002,\ 0.003,\ 0.004~{\rm GeV}^4,\] respectively. Thus, increasing the gluon-condensate parameter delays the disappearance of the $Z_c$ ground-state peak. This trend is qualitatively similar to that found in the $\pi_1$ channel, although the $Z_c$ peak position is much less sensitive to $c$. For the $Z_c$ channel, this increase of $T_{\rm diss}^{Z_c}$ cannot be attributed to a sizable upward shift of the vacuum mass, since $M_{Z_c,0}^{\rm vac}(c)$ changes only weakly over the same range of $c$. It rather reflects the modification of the finite-temperature spectral shape by the gluon-condensate geometry.

\begin{table}[t]
\centering
\small
\setlength{\tabcolsep}{3.2pt}
\renewcommand{\arraystretch}{1.45}

\resizebox{\textwidth}{!}{%
\begin{tabular}{
C{1.30cm}
C{1.30cm}
C{1.75cm}
C{1.55cm}
C{1.45cm}
C{1.75cm}
C{1.75cm}
C{1.55cm}
C{1.45cm}
C{1.75cm}
}
\hline
&
&
\multicolumn{4}{c}{$\pi_1$ channel}
&
\multicolumn{4}{c}{$Z_c$ channel}
\\
\cline{3-6}
\cline{7-10}

$c$
\newline $({\rm GeV}^{4})$
&
$T_c$
\newline $({\rm MeV})$
&
$M_{\pi_1,0}^{\rm vac}$
\newline $({\rm MeV})$
&
$T_{\rm diss}^{\pi_1}$
\newline $({\rm MeV})$
&
$\dfrac{T_{\rm diss}^{\pi_1}}{T_c}$
&
$\dfrac{T_{\rm diss}^{\pi_1}}
{M_{\pi_1,0}^{\rm vac}}$
&
$M_{Z_c,0}^{\rm vac}$
\newline $({\rm MeV})$
&
$T_{\rm diss}^{Z_c}$
\newline $({\rm MeV})$
&
$\dfrac{T_{\rm diss}^{Z_c}}{T_c}$
&
$\dfrac{T_{\rm diss}^{Z_c}}
{M_{Z_c,0}^{\rm vac}}$
\\
\hline

0.001
& 108.9
& 1393.7
& 51
& 0.468
& 0.0366
& 3889.6
& 48
& 0.441
& 0.0123
\\

0.002
& 129.5
& 1469.7
& 65
& 0.502
& 0.0442
& 3889.3
& 61
& 0.471
& 0.0157
\\

0.003
& 143.4
& 1535.4
& 75
& 0.523
& 0.0488
& 3892.3
& 72
& 0.502
& 0.0185
\\

0.004
& 154.0
& 1591.2
& 83
& 0.539
& 0.0522
& 3899.0
& 82
& 0.532
& 0.0210
\\
\hline
\end{tabular}%
}

\caption{Ground-state dissociation temperatures of the $\pi_1$ and $Z_c$ mesons for different values of $c$.}
\label{tab:dissociation-temperatures}
\end{table}

The comparison with the $\pi_1$ channel shows a clear channel dependence. The $\pi_1$ ground-state peak is more sensitive to the gluon-condensate parameter in its peak position, whereas the $Z_c$ ground-state peak is much more stable in frequency. However, according to the dissociation criterion, the $Z_c$ ground state disappears at lower temperatures than the $\pi_1$ ground state for the same values of $c$. Therefore, the stability of the peak position should be distinguished from the thermal survival of the resonance state. An opposite conclusion has been obtained in~\cite{Toniato:2025gts}, where hybrid mesons  are found to dissociate at a lower temperature than tetraquark states. This discrepancy may be due to the absence of a bulk Proca mass term in their probe action.

As shown in table~\ref{tab:dissociation-temperatures}, the dissociation temperatures remain below the corresponding Hawking--Page transition temperatures for all benchmark values of $c$. In particular, the ratios $T_{\rm diss}^{\pi_1}/T_c$ vary between $0.468$ and $0.539$, whereas $T_{\rm diss}^{Z_c}/T_c$ remain in the narrower range
$0.441$--$0.532$. Therefore, the increase of dissociation temperature with $c$ follows the increase of the characteristic background scale $T_c\propto c^{1/4}$.

A complementary comparison is provided by normalizing the dissociation temperature to the corresponding zero-temperature ground-state mass. For the $\pi_1$ channel,
$T_{\rm diss}^{\pi_1}/M_{\pi_1,0}^{\rm vac}$ increases from $0.0366$ to $0.0522$ as $c$ is varied from $0.001$ to $0.004~{\rm GeV}^{4}$. The result of lower condensate is basically consistent with those obtained in~\cite{Bellantuono:2014lra}, which reports a dimensionless dissociation temperature $t=0.1$, where $t\equiv 2\sqrt{2}\,T_{\rm diss}^{\pi_1}/M_{\pi_1,0}^{\rm vac}$. For the $Z_c$ channel, the
corresponding ratio increases from $0.0123$ to $0.0210$. The substantially smaller values in the $Z_c$ channel reflect both its larger vacuum mass scale and its lower dissociation temperature. Within each channel, the increase of $T_{\rm diss}^{X}/M_{X,0}^{\rm vac}$ indicates that the growth of the dissociation temperature with $c$ cannot be attributed solely to the $c$ dependence of the vacuum mass. This is particularly evident for the $Z_c$ channel, whose ground-state vacuum mass changes only mildly, while its dissociation temperature increases appreciably.

\section{Summary and discussion}\label{sec:summary}

In this work, we have developed a holographic framework for computing retarded correlators and spectral functions of exotic spin--1 mesons with an explicit bulk mass term. The inclusion of the Proca mass is important for exotic channels, since it keeps track of the canonical ultraviolet dimension of the boundary interpolating operator. In this way, hybrid and tetraquark-like spin--1 channels can be treated within a unified five-dimensional effective description, while still retaining their different ultraviolet scaling
properties.

We first derived the membrane-flow equations for a massive vector fluctuation at zero spatial momentum propagating in a general diagonal black-hole geometry. The spectral function can be obtained from the boundary value of the membrane conductivity. For a massive bulk vector field, the near-boundary behavior differs from the Maxwell case. Therefore, the holographic source normalization requires an additional cutoff-dependent factor. We also introduced the spectral function $\widehat{\rho}_X(\omega)$, which removes the leading ultraviolet power-law growth and provides a convenient dimensionless quantity for displaying the thermal evolution of resonance peaks.

We then applied this formalism to two exotic spin--1 channels, namely the hybrid $\pi_1$ channel and the tetraquark-like $Z_c$ channel, in a gluon-condensate background. The channel-dependent soft-wall profile $\Phi_X(z)$ was kept fixed for each channel. Thus, the gluon-condensate parameter $c$ enters the calculation through the background geometry rather than through a refitting of the channel profile. This separation is useful for identifying the direct effect of the gluon-condensate deformation on the
vacuum spectrum and on the finite-temperature spectral functions.

At zero temperature, the finite-$c$ background modifies the normal-mode spectrum. The upper bound was selected using two complementary considerations. Its upper endpoint, $c=0.004~\mathrm{GeV}^{4}$, is motivated by the
quantitative matching to the phenomenological gluon condensate discussed in Sec.~\ref{subsec:gluon-background}. For $\alpha_s=0.73$, this value corresponds to
$\langle(\alpha_s/\pi)G^2\rangle\simeq0.0116~{\rm GeV}^4$, in good agreement with the commonly used QCD sum-rule value
$0.012~{\rm GeV}^4$. The same value also yields a Hawking--Page
transition temperature of approximately $154~{\rm MeV}$, consistent with the transition scale inferred from lattice calculations and relativistic heavy-ion phenomenology. In addition, the ground-state vacuum-mass shifts remain within the adopted $20\%$ tolerance throughout the selected range. The selected range therefore provides a phenomenologically motivated deformation scale while preserving the continuous identification of the vacuum ground-state channels. In the $\pi_1$ channel, the masses increase monotonically with $c$, and the excited states are more sensitive to the background deformation than the ground state. The relative
ground-state mass shifts are approximately $2.9\%$, $8.5\%$, $13.3\%$, and $17.5\%$ for $c=0.001$, $0.002$, $0.003$, and $0.004~{\rm GeV}^4$, respectively.  In contrast, the $Z_c$ ground-state mass is much less sensitive to $c$, with relative shifts below $0.3\%$ over the same range. This shows that the $\pi_1$ channel is more sensitive to the gluon-condensate deformation in its vacuum mass spectrum, whereas the $Z_c$ ground-state mass is comparatively stable.

At finite temperature, the discrete normal-mode spectrum is replaced by the real-frequency spectral function obtained from the retarded correlator with incoming-wave boundary conditions at the black-hole horizon. For each value of $c$, the thermal peak is tracked from the corresponding $c$-dependent low-temperature ground-state peak. The dissociation temperature is defined operationally as the temperature at which this ground-state peak ceases to be a well-defined isolated maximum.

For the $\pi_1$ channel, increasing temperature shifts the ground-state peak toward lower frequency, reduces the peak height, and increases the thermal width. Increasing the gluon-condensate parameter, on the other hand, moves the low-temperature ground-state peak to higher frequency, consistent with the upward shift of the corresponding vacuum mass. Using the normalized-height criterion, we find $T_{\rm diss}^{\pi_1}=51,\ 65,\ 75,\ 83~{\rm MeV}$ for $c=0.001,\ 0.002,\ 0.003,\ 0.004~{\rm GeV}^4$, respectively. Thus, larger values of $c$ delay the disappearance of the $\pi_1$ ground-state peak in absolute temperature. This behavior reflects the combined effect of the gluon-condensate deformation on the spectral shape and on the vacuum mass scale.

For the $Z_c$ channel, the peak position is much less sensitive to the gluon-condensate parameter, in agreement with the weak $c$ dependence of the zero-temperature ground-state mass. Nevertheless, the spectral shape is still modified by the background deformation. The corresponding dissociation temperatures are found to be $T_{\rm diss}^{Z_c}=48,\ 61,\ 72,\ 82~{\rm MeV}$ for $c=0.001,\ 0.002,\ 0.003,\ 0.004~{\rm GeV}^4$, respectively. Therefore, the gluon-condensate parameter also delays the disappearance of the $Z_c$ resonance state within the same criterion. However, for all values of $c$ considered here, the $Z_c$ ground-state peak melts at a lower temperature than the corresponding $\pi_1$ peak. This demonstrates that the stability of the peak position should not be confused with the thermal survival of the resonance structure. An opposite conclusion has been obtained in~\cite{Toniato:2025gts}, where hybrid mesons  are found to dissociate at a lower temperature than tetraquark states. This discrepancy may be due to the absence of a bulk Proca mass term in their probe action. Unlike the $\pi_1$ channel, this increase of $T_{\rm diss}^{Z_c}$ is not driven by a large vacuum mass shift. It is instead associated with the modification of the thermal spectral shape induced by the gluon-condensate geometry.

Overall, our results show that the gluon-condensate background has a nontrivial and channel-dependent impact on exotic spin--1 spectral functions. The $\pi_1$ channel exhibits a stronger $c$-dependent shift of the vacuum and low-temperature peak positions, while the $Z_c$ channel remains more stable in frequency but has a lower dissociation temperature. In both channels, the increase of $c$ raises the dissociation temperature obtained from the dissociation criterion. This indicates that the gluon-condensate background does not simply enhance melting; rather, it changes both the vacuum mass scale and the finite-temperature spectral shape.

Several extensions of the present work are possible. First, the thermal masses and FWHM widths used here are effective real-frequency observables. A more direct pole-based characterization of thermal masses and widths would require solving the corresponding quasinormal-mode problem in the complex frequency plane. Second, the present analysis keeps the channel-dependent soft-wall profile fixed as $c$ is varied. A refitting of the profile parameters in the deformed background could provide an alternative scheme for separating vacuum spectral calibration from medium effects. Finally, it would be interesting to extend the same massive-vector membrane-flow framework to other exotic
channels and to backgrounds with additional external parameters, such as chemical potential, magnetic field, rotation, or anisotropy.

\section*{Acknowledgments}
Zhou-Run Zhu is supported by the Natural Science Foundation of Henan Province of China under Grant No. 242300420947. Zhou-Run Zhu is also supported by the High Level Talents Research and Startup Foundation Projects for Doctors of Zhoukou Normal University No. ZKNUC2023018. Yan-Qing Zhao is supported by the National Natural Science Foundation of China (NSFC) under Grant No. 12505151.

\bibliographystyle{utphys}
\bibliography{ref}

\end{document}